\documentclass[aps,prl,reprint,superscriptaddress,amsmath,amssymb]{revtex4-2}

\usepackage{graphicx}
\begin{document}

\title{Comment on “Thermal, Quantum Antibunching and Lasing Thresholds from Single Emitters to Macroscopic Devices”}

\author{Andrey~A.~Vyshnevyy}
\email[]{andrey.vyshnevyy@phystech.edu}
\author{Dmitry~Yu.~Fedyanin}

\affiliation{Laboratory of Nanooptics and Plasmonics, Center for Photonics and 2D Materials, Moscow Institute of Physics and Technology, Dolgoprudny 141700, Russian Federation}

\date{\today}

\begin{abstract}
\end{abstract}

\maketitle

In a recent Letter \cite{Carroll2021-jx}, M. A. Carroll et al. derived a model to analytically determine regimes of thermal, collective anti-bunching, and laser emission for emitters in a cavity. According to their model, nanolasers exhibit a distinct threshold at which the coherent laser field $\langle b\rangle$ emerges from a bifurcation. The amplitude of this coherent field increases with a further increase in the pump rate. Such a behavior contrasts the usual view of the transition to lasing in single-mode high-$\beta$ nanolasers, according to which strong spontaneous emission into the lasing mode results in a smooth transformation from the thermal state to the coherent state as the pump rate increases. Furthermore, as follows from Eqs.~(4), (5) in Ref.~\cite{Carroll2021-jx}, and Eq.~(16) in Supplemental Material, the equation of motion for this coherent field above the bifurcation point is $d_t \langle b \rangle = -i\Omega \langle b \rangle$, which implies that the emitted radiation is ideally monochromatic. Here, we demonstrate that the authors have ignored important terms in the equations of their model, which caused the bifurcation and the ideally monochromatic field to emerge at a finite pump rate. 

Without anomalous averages $\langle b\rangle,\, \langle v^\dagger c\rangle$ and their conjugates, i.e., below the bifurcation point predicted in Ref.~\cite{Carroll2021-jx}, the equation of motion for the photon-assisted polarization, corresponding to Eq.~(2) in Ref.~\cite{Carroll2021-jx}, should read as	
\begin{multline}
d_t \delta\langle bc_l^\dagger v_l\rangle = -(\gamma_c+\gamma+i\Delta\nu)\langle bc_l^\dagger v_l\rangle + g^*\left[\langle c_l^\dagger c_l\rangle \right. \\ 
\left. + \delta\langle b^\dagger b\rangle (2\langle c_l^\dagger c_l\rangle-1\rangle) \right] + \mbox{\boldmath${g^*\delta\langle b^\dagger bc_l^\dagger c_l\rangle}$}\\ 
 - \mbox{\boldmath${g^*\delta\langle b^\dagger bv_l^\dagger v_l\rangle + g^*\sum_{m\ne l}\delta\langle c_l^\dagger v_l v_m^\dagger c_m\rangle}$}. \label{eq:correct_eq}
\end{multline}
The last three terms in this equation were omitted in Eq.~(2) in Ref.~\cite{Carroll2021-jx} without justification. However, nothing can justify the omission of the last term, which is related to the dipole moment of the gain medium. To demonstrate this, we derive the semiclassical analog of Eq.~(\ref{eq:correct_eq}) from the Maxwell-Bloch equations
\begin{eqnarray}
d_t b &&= -(i\nu+\gamma_c)b+g^*P,\\ 
d_t P &&= -(i\nu_\varepsilon+\gamma)P+Ngb(\langle c^\dagger c\rangle-\langle v^\dagger v\rangle), 
\end{eqnarray}
where $P$ is the total electric dipole moment of the gain medium and other notations are the same as in Ref.~\cite{Carroll2021-jx}, and obtain 
\begin{eqnarray}
d_t(bP^*) &&= -(\gamma_c+\gamma+i\Delta \nu)bP^* \nonumber \\ 
&&+Ng^*|b|^2(\langle c^\dagger c\rangle-\langle v^\dagger v\rangle)+g^*|P|^2.\label{eq:semiclass}
\end{eqnarray}

The term-by-term comparison between Eqs.~(\ref{eq:correct_eq}) and (\ref{eq:semiclass}) shows that $bP^*$ represents $\sum_l\langle bc^\dagger_l v_l\rangle$, and $|P|^2$ is the classical equivalent of $\sum_{m,l}\langle c_l^\dagger v_l v_m^\dagger c_m\rangle$. The terms without classical counterparts represent spontaneous emission and correlations arising due to the discrete nature of photons and electrons. The authors of the Letter \cite{Carroll2021-jx} include the dipole moment $P$ only via the anomalous average $\langle v^\dagger c\rangle$, thereby completely discarding it below the bifurcation point. Thus, by omitting $\sum_{m\ne l}\delta\langle c_l^\dagger v_l v_m^\dagger c_m\rangle,$ they produced a quantum model that inherently breaks the quantum-classical correspondence.

It has already been shown that the collective behavior of emitters affects the total emission rates and the statistics of emitted photons \cite{Leymann2015-pb}. The recent work \cite{Vyshnevyy2021-oq} demonstrates that the macroscopic dipole moment of the gain medium induces polaritonic transformation of the cavity mode, which, as in Ref.~\cite{Leymann2015-pb}, modifies the spontaneous and stimulated emission rates. As a result, the population $\langle c^\dagger c\rangle_\mathrm{flc}$  required for the full compensation of the cavity loss by stimulated emission  \textit{exactly} matches the bifurcation population $\langle c^\dagger c \rangle_\text{th}$ in the Maxwell-Bloch equations discussed in Ref.~\cite{Carroll2021-jx}. The equation for the steady-state photon number has the following form:
\begin{equation}
0=d_t \langle b^\dagger b\rangle =R_\mathrm{spont}+(G-2\gamma_c)\langle b^\dagger b\rangle, 
\end{equation}
where $R_\mathrm{spont}$ is the spontaneous emission rate and $G\langle b^\dagger b\rangle$ is the stimulated emission rate, while $2\gamma_c$ is the photon loss rate in the cavity mode. Thus, at any pump rate, $G<2\gamma_c$, $\langle c^\dagger c\rangle< \langle c^\dagger c \rangle_\mathrm{flc}=\langle c^\dagger c\rangle_\mathrm{th}$, and, therefore, the bifurcation point is beyond reach.
By contrast, without the term $\sum_m\delta\langle c_l^\dagger v_l v_m^\dagger c_m \rangle$, Eqs. (2) and (3) of Ref.~\cite{Carroll2021-jx} predict full compensation of the cavity loss by the stimulated emission at
\begin{equation}
\langle c^\dagger c \rangle_\mathrm{flc}=\frac{1}{2}+\frac{\gamma_c(\gamma_c+\gamma)}{2N|g|^2}\left[1+\left(\frac{\Delta\nu}{\gamma_c+\gamma}\right)^2\right]>\langle c^\dagger c \rangle_\text{th},    
\end{equation}
which puts the bifurcation point within reach. Note, usually $\gamma_c\ll \gamma$, and, therefore, the polarization correlation term $\sum_{m\ne l}\delta\langle c_l^\dagger v_l v_m^\dagger c_m\rangle$ is very small, resulting in an equally small correction to $\langle c^\dagger c\rangle_\mathrm{flc}$ when properly accounted for. However, this small correction is still sufficient to make the bifurcation point unattainable. 

To sum up, in the nanolaser model proposed in Ref.~\cite{Carroll2021-jx}, the bifurcation and abrupt transition to lasing at a finite pump rate, even in high-$\beta$ nanolasers, were obtained because of the unjustified truncation of the dipole moment of the gain medium from the equations of motion. The proper account of the truncated terms within the studied Janes-Cummings quantum model makes the predicted bifurcation point unattainable.

%

\end{document}